\documentclass[12pt]{article}
\usepackage{epsf,epsfig,psfig}
\newcommand{\be}{\begin{equation}}
\newcommand{\ee}{\end{equation}}
\def\slashed{{/}\mskip-11.0mu}

\begin{document}
\begin{titlepage}
\begin{center}
{\Large 
Perturbative renormalization in parton distribution functions 
using Overlap fermions and Symanzik improved gluons
}
\end{center}
\vskip 1.3cm
\centerline{
M. Ioannou and H. Panagopoulos
}

\vskip 0.4cm
\centerline{\sl
Department of Physics, University of Cyprus,
                      Nicosia CY-1678, Cyprus
}
\centerline{\sl
e-mail: {\tt mike80bi@yahoo.gr, haris@ucy.ac.cy}
} 

\vskip 3.cm

\begin{abstract}
We calculate the 1-loop renormalization of the fermion self-energy,
all local fermion bilinears, as well as a set of extended 
bilinears which form a basis corresponding to moments of the parton
distribution functions. 

 We use the overlap action for fermions and Symanzik improved action
 for gluons. 

 Our results are presented as a function of the overlap parameter
 $\rho$ and the parameters entering the Symanzik action.

\end{abstract}

\end{titlepage}

\section{Introduction.}

In recent years, simulations in lattice QCD have employed an ever
increasing number of ``improved'' actions for gluon and fermion
fields. This practice is well motivated, since it leads to a reduction
of ${\cal O}(a)$ effects ($a$: lattice size), thus resulting in a
faster approach to continuum physics. In the case of fermions, 
certain improved actions carry the crucial added advantage of preserving
chiral symmetry on the lattice.

The variety of lattice actions presently in use calls for the
computation of a large number of renormalization functions,
before a comparison to physical results can be attempted; such
functions often pertain to fermion composite operators, whose matrix
elements are widely used to extract mass spectra, decay rates and
transition amplitudes.

A set of operators which are of particular interest are fermion
bilinears; they can be classified according to the representations of
the rotation group, and the global flavor group. Operators of this
kind appear naturally in the operator product expansion (OPE), as
applied to hadronic physics; in the Bjorken limit, involved in studies
of deep inelastic scattering, the OPE for a product of hadronic
currents takes the form:
\be
J(x)J(0)\sim\sum_{n,i}C^{n,i}(x^2)x^{\mu_1}...x^{\mu_n}O_{\mu_1...\mu_n}^{(n,i)}(0)
\ee

The forward matrix elements of the local operators $O^{(n,i)}$
   appearing in this expansion are directly related to the moments of
   hadron structure functions. 
   The dominant contribution in the expansion is given by operators
   whose twist (dimension minus spin) equals two, which in the flavor
   non-singlet case means the symmetric traceless operators \cite{ref1} 
\be
O_{\mu\mu_1...\mu_n}=\bar{\psi}\gamma_{\{\mu} D_{\mu_1}...D_{\mu_n\}}\frac{\lambda^a}{2}\psi
\ee
\be
O_{\mu\mu_1...\mu_n}^{(5)}=\bar{\psi}\gamma_{\{\mu}\gamma_5 D_{\mu_1}...D_{\mu_n\}}\frac{\lambda^a}{2}\psi
\ee
where $\lambda^a$ are flavor matrices, and the curly brackets denote
   symmetrization over Lorentz indices.
 A proper renormalization is
   required for these operators, before one can relate their matrix elements,
   as extracted from numerical simulations, to physically observable
   quantities. 

In this work we employ Symanzik improved actions for gluons. We
consider, in particular, actions made up of closed Wilson loops
with up to 6 links; they are the most widely used cases in numerical
simulations. This class of actions is described by 3 independent
parameters which are tuned either in perturbation theory (at tree
level or beyond) or through simulation, so as to minimize 
${\cal O}(a)$ effects. Particular choices of action within this class, aside
from the standard Wilson action, are: tree level Symanzik improved,
``tadpole improved'' L\"uscher-Weisz, Iwasaki and DBW2.

For the fermion part of the action we employ
Neuberger's overlap-Dirac operator. This action is characterized by
one free parameter. The definition of this operator involves the
inverse square root of $X^{\dag}X$, where $X$ is the Wilson-Dirac
operator with a negative mass term; this operation is extremely costly
in numerical simulations, and its viability depends critically on the
condition number of $X^{\dag}X$. Thus, an improved action for gluons becomes
all the more important, since it tends to decrease the condition
number~\cite{DHK}.

The quantities whose renormalizations we set out to compute, to 1-loop
order in lattice perturbation theory, are:
\begin{itemize}
\item[a)] The quark propagator
\be
\langle\psi_\alpha(x)\,{\bar\psi}_\beta(y)\rangle
\ee
\item[b)] Local fermion bilinears.
\be
O_X=\bar{\psi}(x)\Gamma^X\psi(x)
\ee
\item[c)] Operators which measure the first moment of quark momentum
  distributions. 
\be
O_{V_2}=\bar{\psi}\gamma_{\{1}D_{4\}}\psi
\ee
\be
{O'}_{V_2}=\bar{\psi}\gamma_4 D_4
  \psi-\frac{1}{3}\sum_{i=1}^3\bar{\psi}\gamma_i D_i\psi 
\ee
\item[d)] Operators which measure the first moment of quark helicity
  distributions. 
\be
O_{A_2}=\bar{\psi}\gamma_{\{1}\gamma_5D_{4\}}\psi
\ee
\be
{O'}_{A_2}=\bar{\psi}\gamma_4\gamma_5 D_4
  \psi-\frac{1}{3}\sum_{i=1}^3\bar{\psi}\gamma_i\gamma_5 D_i\psi 
\ee
\end{itemize}

The symmetrized, covariant derivatives $D$ are defined as: 
$D = \overrightarrow D - \overleftarrow D$, with:
\be  
\overrightarrow D_\mu\,\psi(x) = {1\over 2a} \,
\Bigl[U(x,\mu)\,\psi(x+a\hat\mu) -
  U^\dagger(x-a\hat\mu,\mu)\,\psi(x-a\hat\mu)\Bigr] 
\ee
and similarly for the left derivative $\overleftarrow D$.

Our calculations provide a cross check and an extension of results appearing in
Refs. \cite{ref10,ref3,ref11}.
Work is in progress using operators measuring the second moment of
quark momentum and helicity distributions.

We have performed our calculations for a rather wide selection of
values for the parameters entering the gluon and fermion actions,
corresponding to present and future simulations.
In particular, our results scan the whole range of admissible values for
the overlap parameter. Results for different choices of parameter
values can be provided by the authors upon request.

A synopsis of some of our results can be found in Ref. \cite{ref2}.

\vskip-0.3cm
\section{Calculational setup.}
We denote the lattice action by
\be
S=S_G+S_F
\ee
where $S_G$  is the gluon action, and $S_F$  is the fermion action.
The gluon action we consider is written in standard notation:
\begin{eqnarray}
S_G&=&-\frac{1}{g^2}\bigl(c_0\sum_{plaquette}TrU_{pl}+c_1\sum_{rectangle}TrU_{rtg}\nonumber\\
&&\qquad +c_2\sum_{chair}TrU_{chr}+c_3\sum_{parallelogram}TrU_{plg}\bigr)
\label{action}
\end{eqnarray}
where $U_{pl}$  is the standard plaquette, while the remaining U's
cover all possible closed loops containing up to six links, as
indicated in Fig. 1.

\begin{center}
  \epsfig{file=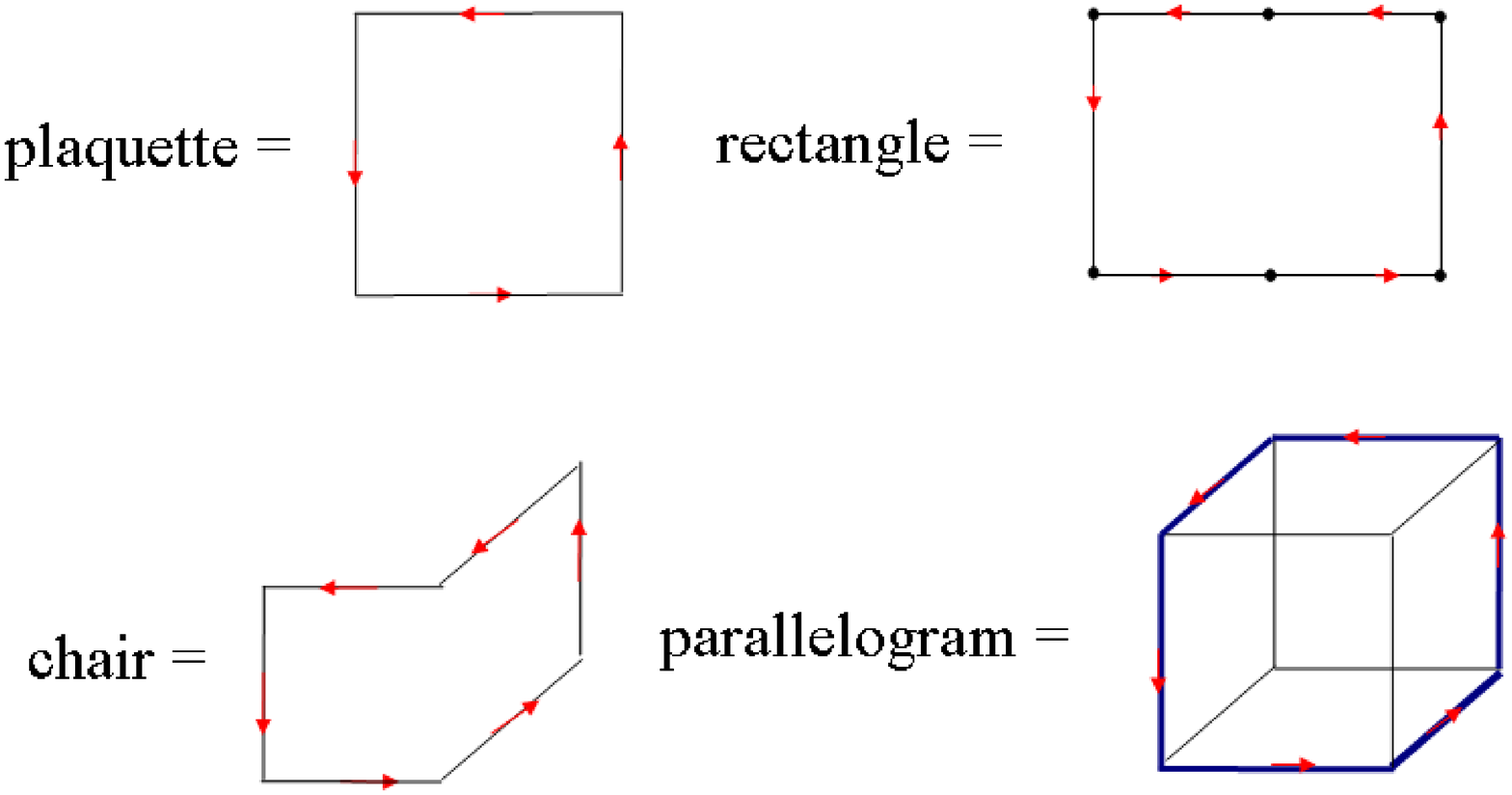, scale=0.36}

\small{Figure 1: Closed loops corresponding to the four terms
  in the gluon action (\ref{action}).} 
\end{center}
The coefficients $c_0,\ c_1,\ c_2,\ c_3$  satisfy the normalization condition:
\be
c_0+8c_1+16c_2+8c_3=1
\ee
As in Ref.\cite{ref3}, we have used the values of $c_0$, $c_1$, $c_2$,
$c_3$, shown in Table 1, where ``Plaquette'' is the standard Wilson
action for gluons, ``Symanzik'' is the  tree-level improved action
\cite{ref4} and TILW is the  tadpole improved Luescher-Weisz action
\cite{ref5,ref6}. Two further choices which we have considered for
$c_i$ are the Iwasaki \cite{ref7} and DBW2 \cite{ref8} actions.
It is worth noting that, at the 1-loop level, these coefficients enter
the calculation only through the propagator and not through any
vertices; given that the propagator does not depend separately on
$c_2$ and $c_3$, but only on their sum, we have considered only
parameter values with $c_2=0$.
\begin{table}[ht]
\begin{center}
\label{values of coefficients}
\begin{tabular}{|c|r@{}l|r@{}l|r@{}l|r@{}l|}
\multicolumn{1}{c}{Action}&
\multicolumn{2}{c}{$c_0$} &
\multicolumn{2}{c}{$c_1$} &
\multicolumn{2}{c}{$c_2$} &
\multicolumn{2}{c}{$c_3$} \\
\hline
Plaquette  &1&.0   &0&.0	        &0&.0	     &0&.0 \\
Symanzik   &1&.66666	&-0&.083333	&0&.0	     &0&.0 \\
TILW1 $\beta=8.60$         &2&.31681     &-0&.151791		&0&.0	     &-0&0128098 \\
TILW2 $\beta=8.45$       &2&.34602         &-0&.154846	&0&.0	     &-0&.0134070 \\
TILW3 $\beta=8.30$        &2&.38698	        &-0&.159128		&0&.0	     &-0&.0142442 \\
TILW4 $\beta=8.20$       &2&.41278	        &-0&.161827	&0&.0	     &-0&.0147710 \\
TILW5 $\beta=8.10$        &2&.44654   &-0&.165353	&0&.0  &-0&.0154645 \\
TILW6 $\beta=8.00$       &2&.48917   &-0&.169805	&0&.0 &-0&.0163414 \\
Iwasaki       &3&.648		&-0&.331	&0&.0	     &0&.0 \\
DBW2        &12&.2688     	&-1&.4086	&0&.0  &0&.0 \\
\hline
\end{tabular}
\end{center}
\caption{The values of coefficients $c_0, c_1, c_2, c_3$\,.}
\end{table}

The action for massless overlap fermions is given by \cite{ref9}
\be
S_F=\sum_{f}\sum_{x,y}\bar{\psi}_x^f D_N(x,y)\psi_y^f
\ee
with
\be
D_N=\rho\,[1+X(X^{\dag}X)^{-\frac{1}{2}}]
\ee
and: $X=D_w-\rho$.  Here $D_w$  is the massless Wilson-Dirac operator with $r = 1$, and $\rho$ is a free parameter whose value must be in the range $0<\rho<2$  in order to guarantee the correct pole structure of $D_N$.
 \section{Results.}
\noindent
{\bf Quark propagator}

Let us consider the massless quark propagator $S_N$  first. In
momentum space, the inverse of $S_N$  can be written as:
\be
S_N^{-1}=i\gamma_{\mu}p_{\mu}(1-\frac{g^2C_F}{16\pi^2}\Sigma_1)
\label{SN}
\ee
with $C_F=(N^2-1)/2N$, and $\Sigma_1(a,p)={\rm log}(a^2p^2)+b_{\Sigma}$
(Feynman gauge). In Eq. (\ref{SN}), the factor in parentheses is the
quark wave function renormalization; in the MOM scheme we have:
\be
Z_\psi^{MOM}(a,\mu) = 1-\frac{g^2C_F}{16\pi^2}\Sigma_1(a,\mu)
\ee
The diagrams that contribute  to one loop order are
shown in Fig. 2. Our results for $b_{\Sigma}$ are shown in Fig. 3
and, on a different scale, in Fig. 4. One observes a divergence of
this quantity for $\rho \to 0$, which can be expected, since the
fermion propagator behaves singularly in this limit. Numerical values
of $b_\Sigma$ are listed in Table 2 for typical values of $\rho$\,: 
$\rho=0.6,\ 1.0,\ 1.4$. 

\begin{center}
\epsfig{file=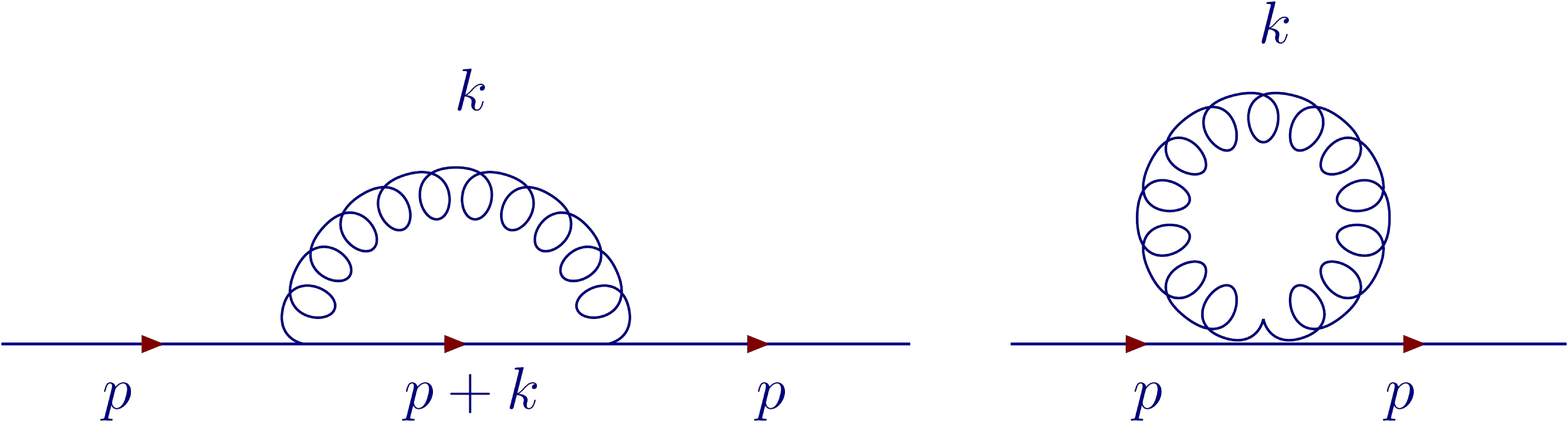, scale=0.10}

\small{Figure 2: One-loop diagrams contributing to the quark
  self-energy.} 
\end{center}

\vspace{3cm}
\begin{center}
\epsfig{file=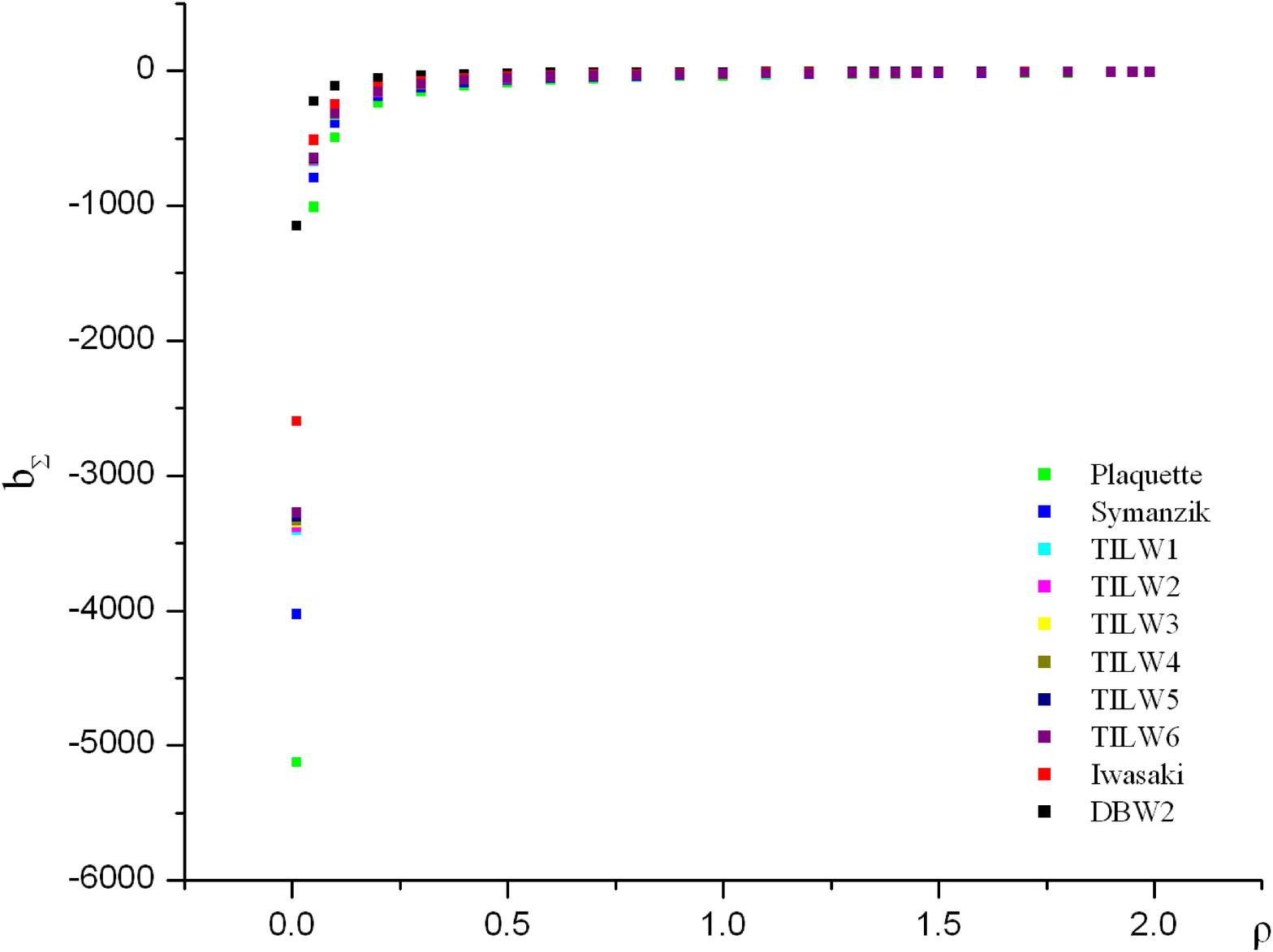, scale=0.22}

\small{Figure 3: $b_{\Sigma}$ as a function of the overlap parameter $\rho$, for different gluon actions.} 
\end{center}
\vspace{-1.5cm}

\begin{center}
\epsfig{file=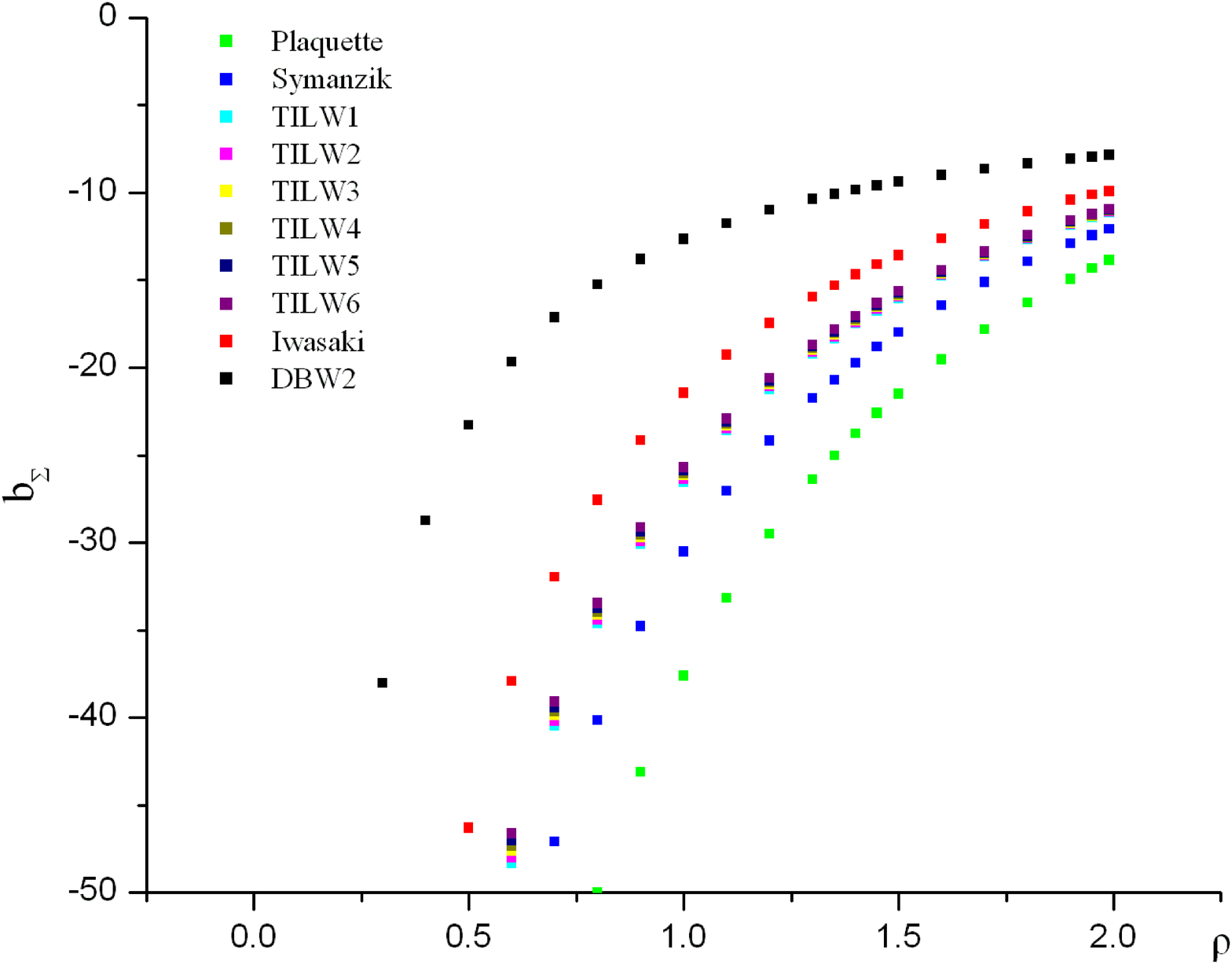, scale=0.22}

\small{Figure 4: $b_{\Sigma}$ as a function of the overlap parameter $\rho$, shown on a different scale.}
\end{center}
\vspace{1cm}

\medskip\noindent
{\bf Quark bilinears}

Let us consider local operators of the form
\be
O_X=\bar{\psi}(x)\Gamma^X\psi(x)
\ee
where $X=S,\ P,\ V,\ A,\ T$, and $\Gamma^S=1,\ \Gamma^P=\gamma_5,\
\Gamma^V=\gamma_{\mu},\ \Gamma^A=\gamma_{\mu}\gamma_5,\
\Gamma^T=\sigma_{\mu\nu}\gamma_5$\,.
We denote the amputated Green's function of the operator $O_X$ by $\Lambda^X.$ 
The final results are (in Feynman gauge):
\be
\Lambda^{S,P}=\{1,\gamma_5\}+\frac{g^2C_F}{16\pi^2}[-4log(a^2p^2)+b_{S,P}]\{1,\gamma_5\}
\ee
\be
\Lambda_{\mu}^{V,A}=\{\gamma_\mu,\gamma_\mu\gamma_5\}+\frac{g^2C_F}{16\pi^2}[-\gamma_\mu(log(a^2p^2)+b_{V,A})+2\frac{p_\mu \slashed {p}}{p^2}]\{1,\gamma_5\}
\ee
\be
\Lambda_{\mu,\nu}^T=\sigma_{\mu,\nu}\gamma_5+\frac{g^2C_F}{16\pi^2}b_T\sigma_{\mu,\nu}\gamma_5
\ee
Where $b_S=b_P$ and $ b_V=b_A$.
Among the diagrams appearing in Fig. 5, only the first one (Vertex)
contributes to $\Lambda^X$. 

\begin{center}
\epsfig{file=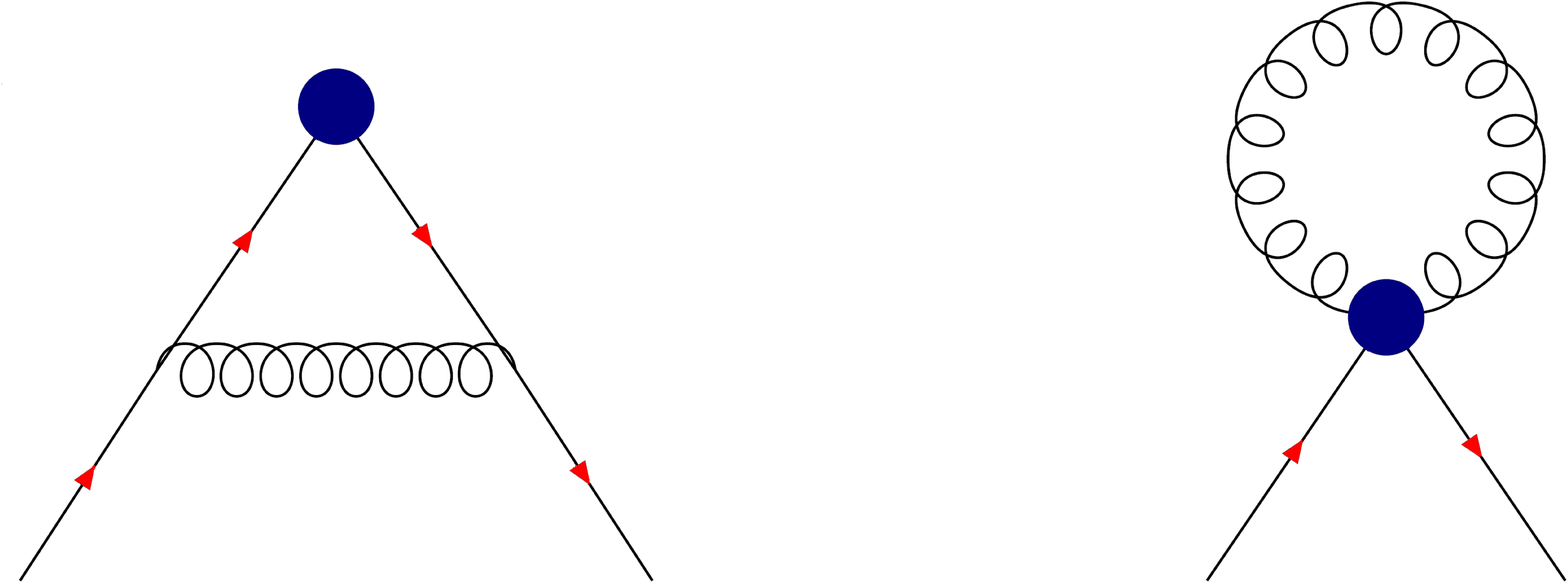, scale=0.08}
\end{center}

\begin{center}
\epsfig{file=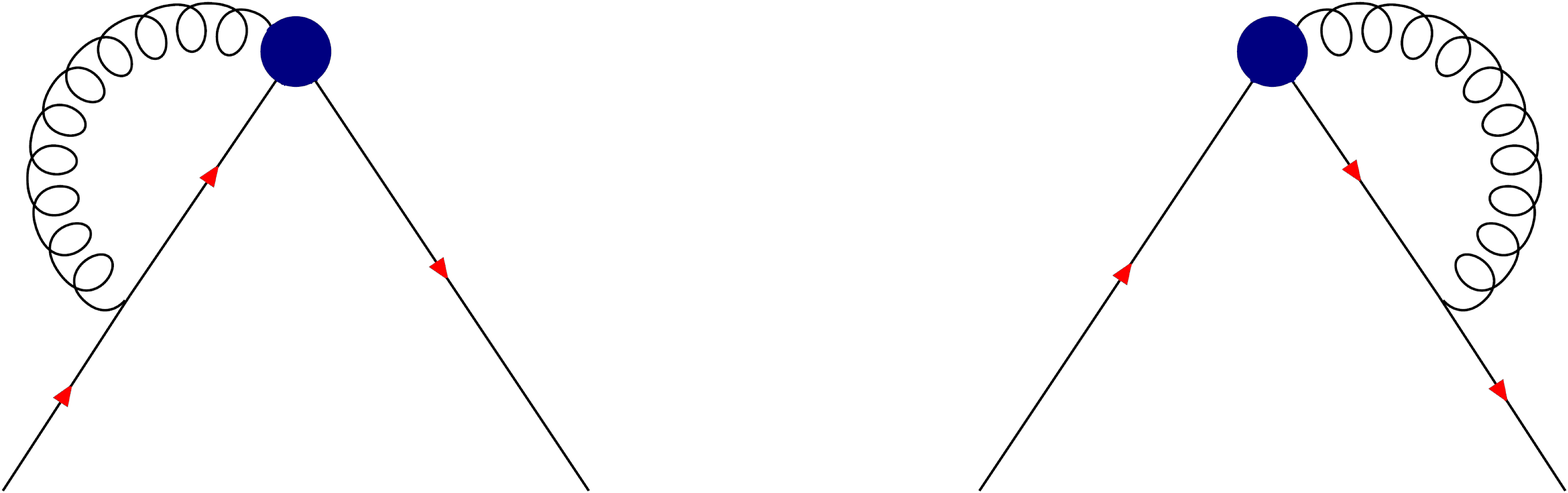, scale=0.08}

\small{Figure 5: One-loop lattice Feynman diagrams contributing to the
  amputated Green's functions of $O_X$ and $O_{\mu\mu_1...\mu_n}.$
  From left to right: vertex, operator tadpole, left sail and right
  sail. }  
\end{center}

Our results 
provide a cross check and an extension of results appearing in
Refs.\cite{ref10,ref3}. They are shown in Figures 3+4 ($b_{\Sigma}$),
6 ($b_{S,P}$), 7 ($b_{V,A}$), 8 ($b_T$) for
typical values of the overlap parameter $\rho$. The same quantities
are also listed in
Tables 2, 3, 4, 5, respectively, for the representative choices:
$\rho=0.6,\ 1.0,\ 1.4$. 
In all cases which can be compared, our results agree with those of
Refs. \cite{ref10,ref3}. There is one exception, regarding the values of
$b_T$  for cases other than the standard plaquette $b_T^{plaq}$ : In
these cases, our results for $b_T-b_T^{plaq}$  have the opposite sign
compared to Ref. \cite{ref3}.\\ 

\vspace{-0.7cm}
\begin{center}
\epsfig{file=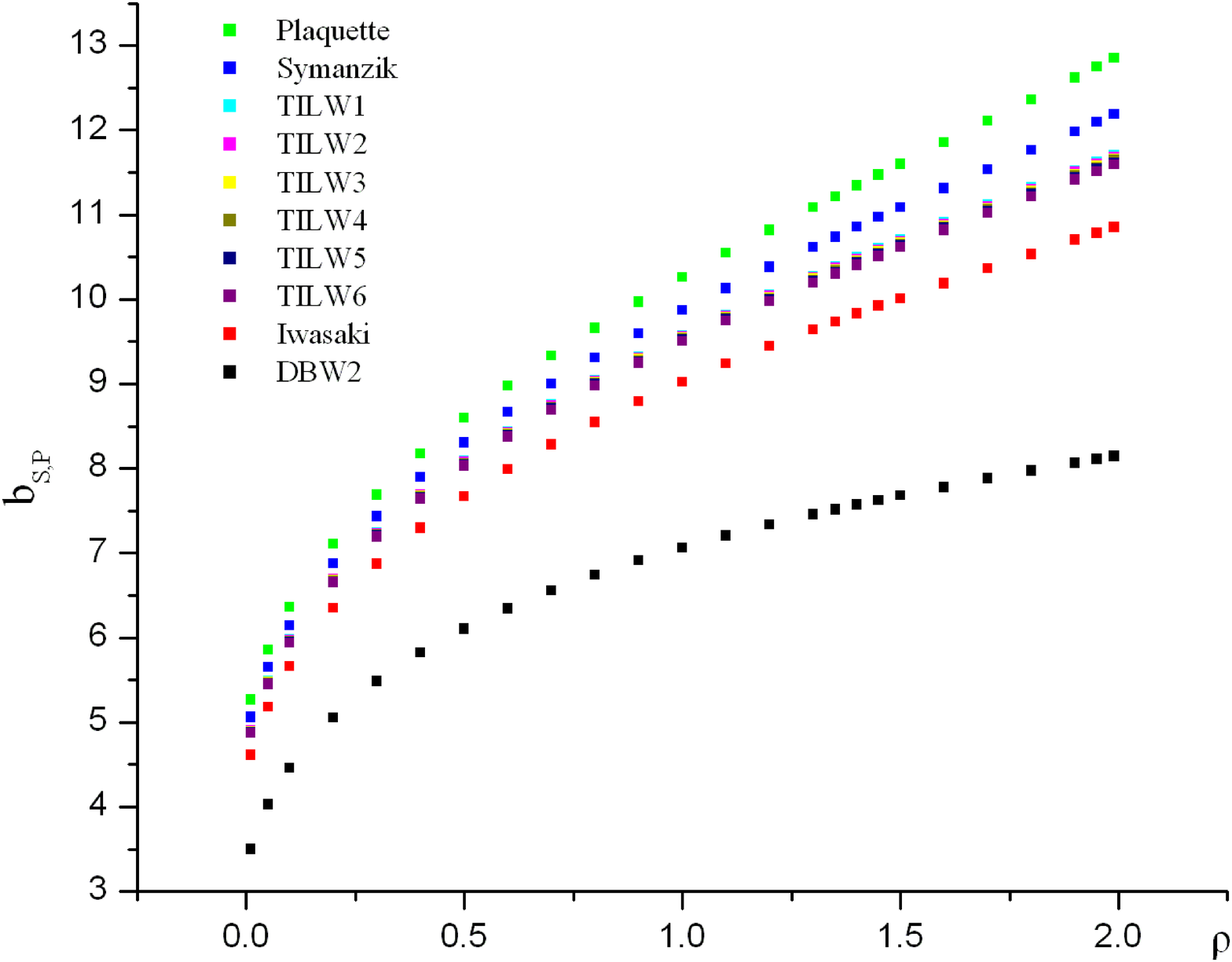, scale=0.22}

\small{Figure 6: $b_{S,P}$ as a function of the overlap parameter $\rho$, for different gluon actions.}
\end{center}
\newpage

\vspace{-1.5cm}
\begin{center}
\epsfig{file=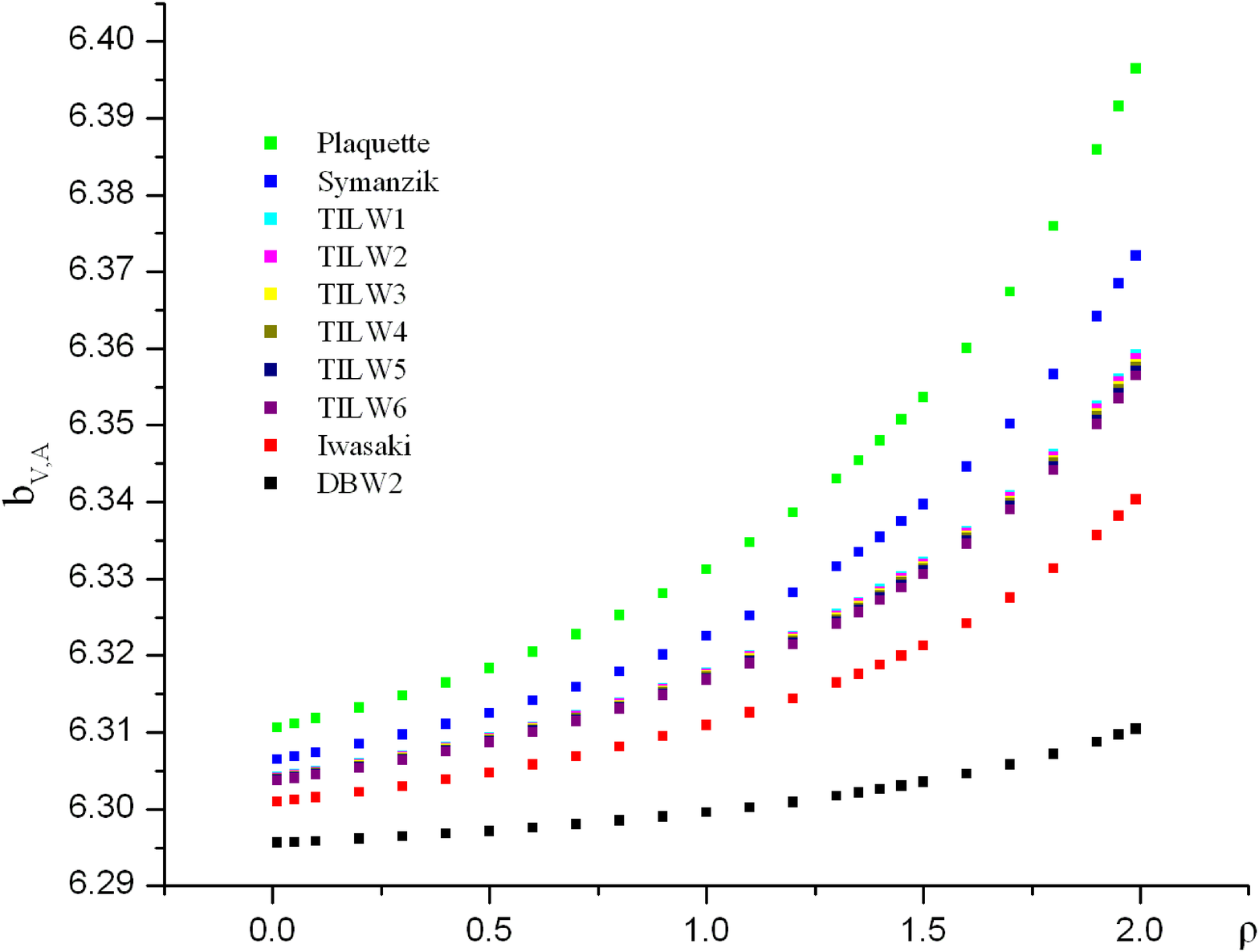, scale=0.22}

\small{Figure 7: $b_{V,A}$ as a function of the overlap parameter $\rho$, for different gluon actions.}
\end{center}
\vspace{-0.8cm}

\begin{center}
\epsfig{file=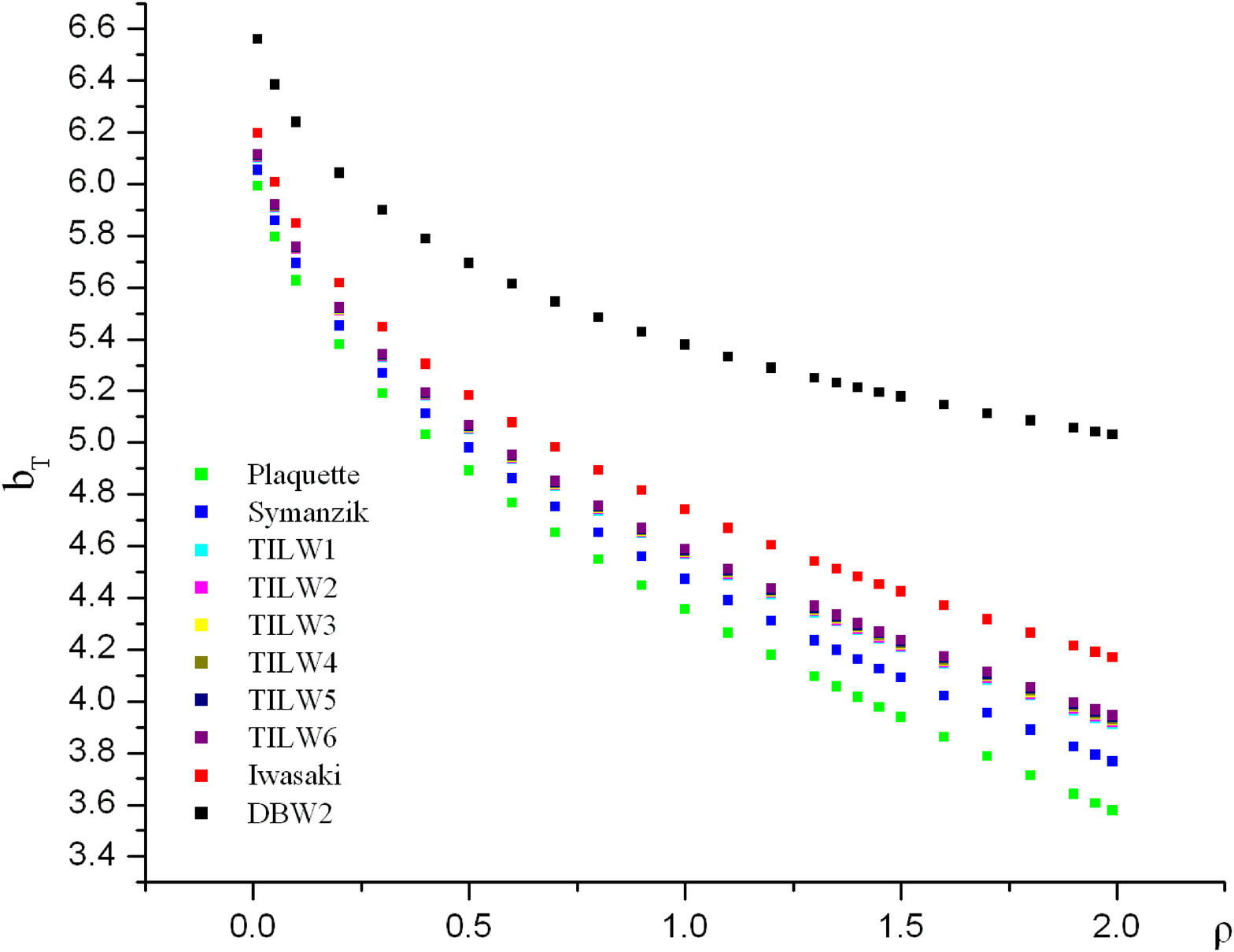, scale=0.22}

\small{Figure 8: $b_{T}$ as a function of the overlap parameter $\rho$, for different gluon actions.}
\end{center}
\newpage

\medskip\noindent
{\bf First moment of quark distributions}.\\

 We have calculated the 1-loop renormalization coefficients of the
 operators $O_{V_2}, O'_{V_2}$  which are the symmetric off-diagonal,
 and the traceless diagonal parts, respectively, of the operator: 
\be
O_{\mu,\nu}=\bar{\psi}\gamma_{\{\mu}D_{\nu\}}\psi
\ee
 The axial counterparts $O_{A_2}, O'_{A_2}$  of the above operators
 renormalize in the same way. Using the notation of Ref. \cite{ref11},
 we find for the amputated Green's function of $O_{\mu,\nu}$ in the
 Feynman gauge: 
\be
\Lambda_{\mu,\nu}(a,p)=\gamma_{\{\mu}p_{\nu\}}+\frac{g^2C_F}{16\pi^2}\,
 \Bigl[(\frac{5}{3}log(a^2p^2)+b)\,\gamma_{\{\mu}p_{\nu\}}+
 b'\,\delta_{\mu\nu}\gamma_{\nu}p_{\nu}-
 \frac{4}{3}\frac{p_{\mu}p_{\nu}}{p^2}\,\slashed    
 {p}\Bigr]\label{Lambdamunu} 
\ee
The quantities $b,\ b'$  correspond to $(b_1+b_2),\  b_4$  of
 Ref. \cite{ref11}, respectively. All diagrams of Fig. 5 contribute in
 this case. 
  The rational coefficients 5/3 and -4/3 in Eq. (\ref{Lambdamunu}) check with
   those of Ref. \cite{ref11}. The values of $b$ and $b'$ are shown in
   Figures 9 and 10; they are in agreement with the quantities
   $b_1+b_2$ and $b_4$, respectively, as defined in Ref. \cite{ref11},
   for the choice of parameters considered in that reference. In
   Tables 6 and 7 we list our results for $b$ and $b'$, respectively,
   with the choice of values for $\rho$\,: $\rho = 0.6,\ 1.0,\ 1.4$.

\begin{center}
\epsfig{file=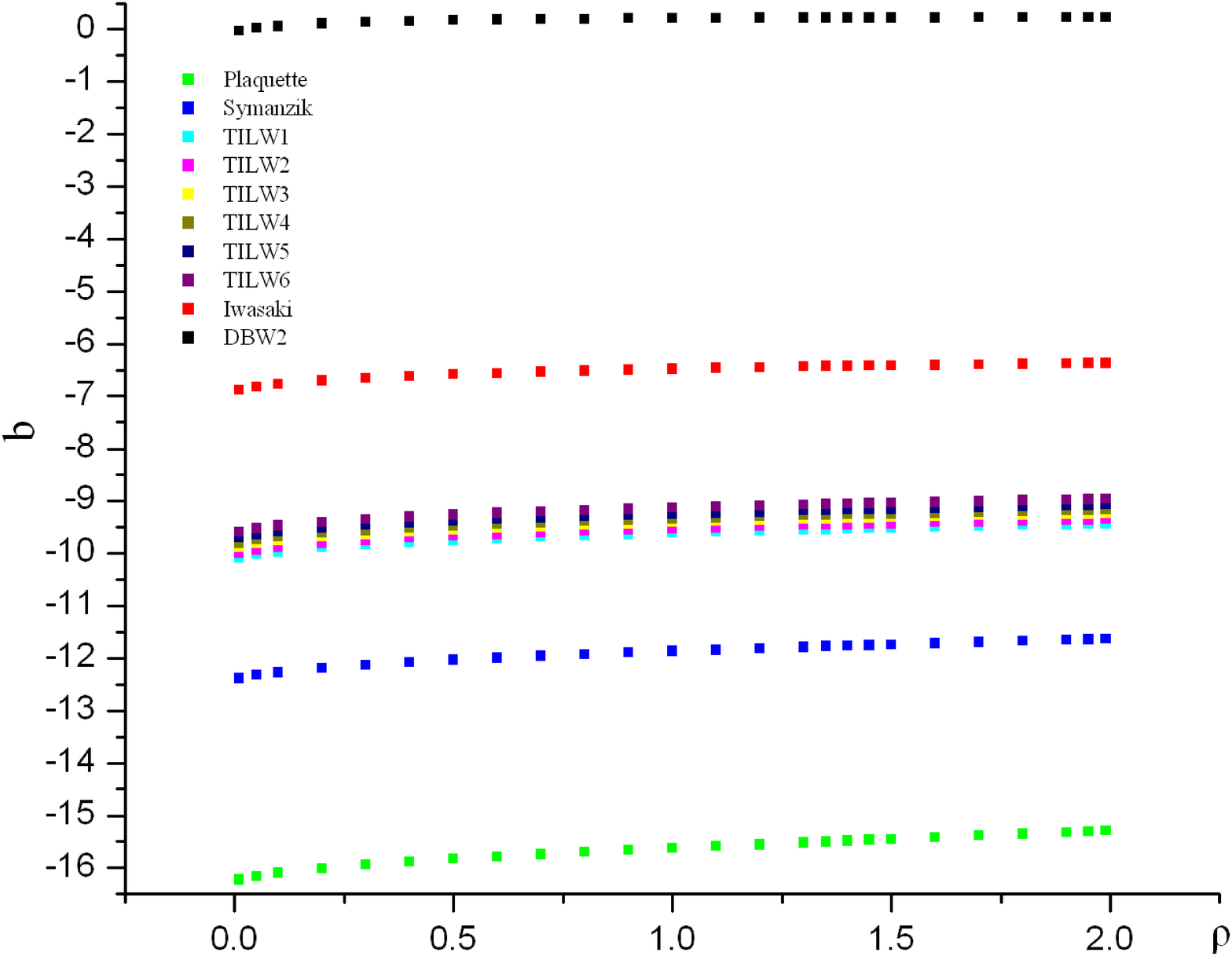, scale=0.22}

\small{Figure 9: $b$ as a function of the overlap parameter $\rho$, for different gluon actions.}
\end{center}

\begin{center}
\epsfig{file=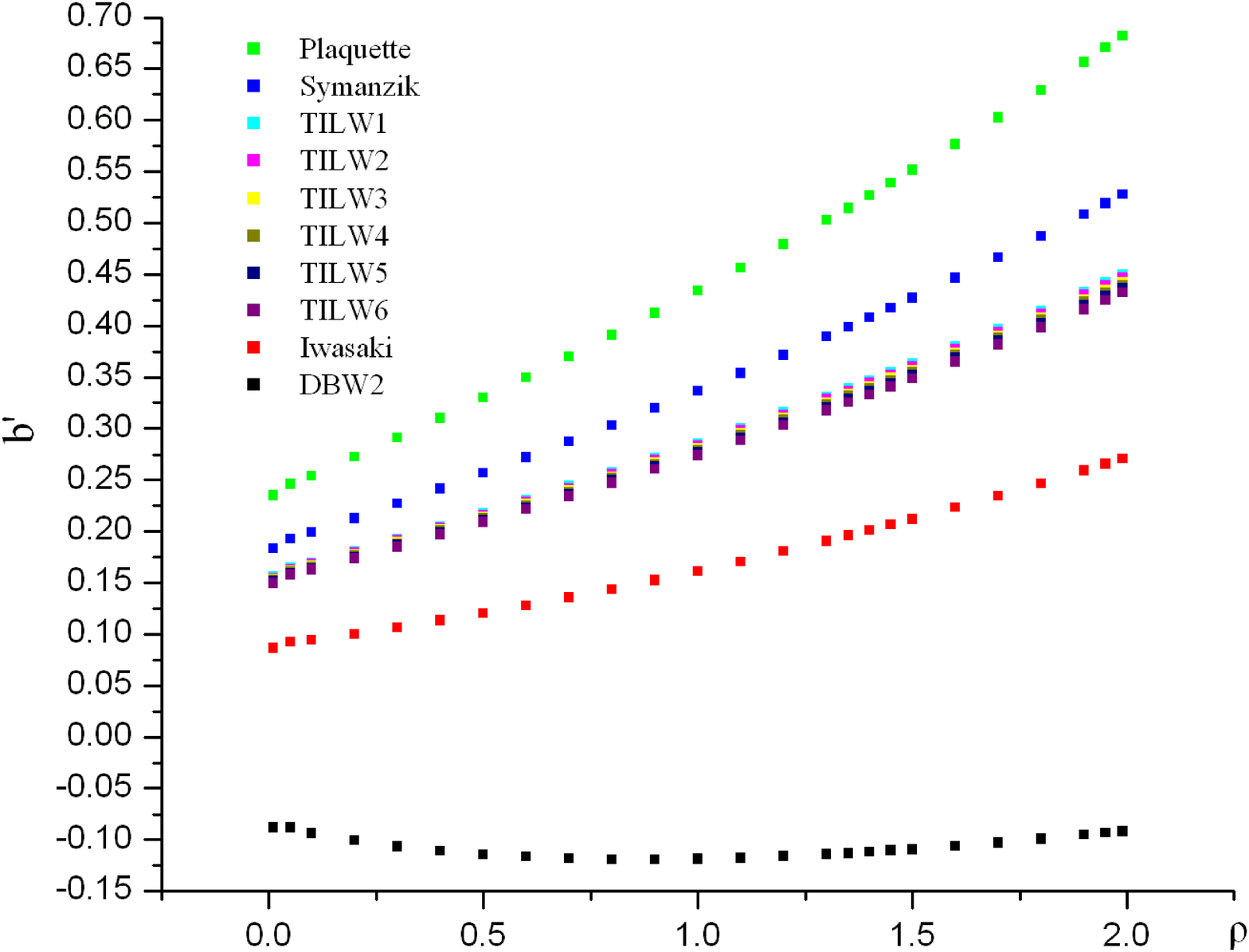, scale=0.22}

\small{Figure 10: $b'$ as a function of the overlap parameter $\rho$, for different gluon actions.}
\end{center}  

All 1-loop calculations presented in this work were performed using
our Mathematica package for lattice perturbation theory. Numerically
inverting the gluon propagator turned out to be less CPU consuming,
as compared to using an analytic expression for the inverse. We used a rather
wide spectrum of values for the Symanzik parameters $c_i$\,; given
that the dependence of 1-loop results on $c_2$ and $c_3$ is only
through their sum $c_2+c_3$\,, we have chosen $c_2=0$ without loss of
generality. For the overlap parameter $\rho$ we have explored in
detail its full spectrum of admissible values. Results for further
values of the 
parameters can be provided by the authors upon request.

We observe a rather smooth dependence of our results on $\rho$, and
this fact allows for a safe interpolation to any additional
$\rho$-values in the allowed range $0<\rho<2$. A possible exception is
the behaviour of $b_\Sigma$ near the endpoint $\rho\to 0$, which can
be expected given the incorrect pole structure of the fermion operator
in that limit. The dependence on $\rho$ is even less pronounced for
the DBW2 action.

Results for the various choices of parameters for the Tadpole Improved
L\"uscher-Weisz (TILW) action are in many cases almost indistinguishable from
one another, due to the proximity in these parameter values;
consequently, it is straightforward to interpolate accurately our
results to other sets of 
values which were not considered in this work.

It is worth noticing that both $b$ and $b'$ are noticeably suppressed
in the case of the DBW2 action. A similar effect has been observed in
the perturbative renormalization of both the topological charge and
susceptibility, using this action~\cite{PS}. This fact is an
indication in favour of this
action for purposes of numerical simulation.

Some related quantities of interest for the study of structure
functions are the second moments of quark momentum and helicity
distributions, as well as non-forward
matrix elements of various fermion bilinears. We expect to address these
issues in a future report.


\newpage
\begin{table}[ht]
\begin{center}
\label{values of bSigma}
\begin{tabular}{|c|r@{}l|r@{}l|r@{}l|}
\multicolumn{7}{c}{{\bf TABLES}}\\
\multicolumn{1}{c}{Action}&
\multicolumn{2}{c}{$\rho=0.6$} &
\multicolumn{2}{c}{$\rho=1.0$} &
\multicolumn{2}{c}{$\rho=1.4$} \\
\hline
Plaquette         &-70&.8442846023(6)  &-37&.6306267984(3) &-23&.7659791950(4)  \\
Symanzik           &-56&.4356742205(6)  &-30&.5049192399(3) &-19&.7319952945(2)   \\
TILW1 $\beta=8.60$ &-48&.315432542(1)  &-26&.5273492659(2) &-17&.5090552592(4)        \\
TILW2 $\beta=8.45$ &-48&.020432621(1)  &-26&.3832716142(5) &-17&.4287897913(1)  \\
TILW3 $\beta=8.30$ &-47&.614837428(1)  &-26&.1852273116(3)  &-17&.3184864381(3)      \\
TILW4 $\beta=8.20$ &-47&.363843382(1)   &-26&.0626992727(4) &-17&.2502579516(4)   \\
TILW5 $\beta=8.10$ &-47&.040843009(1)  &-25&.9050530413(4) &-17&.1624938296(4)   \\
TILW6 $\beta=8.00$ &-46&.641145138(1)   &-25&.7100225495(5) &-17&.0539447038(2)   \\
Iwasaki          &-37&.911348699(2)  &-21&.454735885(1) &-14&.680753624(5)    \\
DBW2             &-19&.673287561(6)   &-12&.675353551(3) &-9&.840831462(1)   \\
\hline
\end{tabular}
\end{center}
\caption{The contribution $b_{\Sigma}$ to the self energy for various actions.}
\begin{center}
\label{values of bS,P}
\begin{tabular}{|c|r@{}l|r@{}l|r@{}l|}
\multicolumn{1}{c}{Action}&
\multicolumn{2}{c}{$\rho=0.6$} &
\multicolumn{2}{c}{$\rho=1.0$} &
\multicolumn{2}{c}{$\rho=1.4$} \\
\hline
Plaquette          &8&.9798281638(5) &10&.26190085663(6) &11&.34335740638(2) \\
Symanzik            &8&.6712333358(5) &9&.87030941138(2) &10&.85599048485(3)\\
TILW1 $\beta=8.60$  &8&.4347849739(6) &9&.57373721242(1)&10&.4927473931(1) \\
TILW2 $\beta=8.45$  &8&.4249104780(5) &9&.56141189336(1) &10&.4777493701(1)\\
TILW3 $\beta=8.30$  &8&.4111618345(5) &9&.54425880565(2) &10&.45688939118(1)\\
TILW4 $\beta=8.20$  &8&.4025516764(5) &9&.53352128912(2) &10&.44383891932(1)\\
TILW5 $\beta=8.10$  &8&.3913573861(6) &9&.51956660551(1) &10&.4268869200(1)\\
TILW6 $\beta=8.00$  &8&.3773207579(2) &9&.50207732589(2) &10&.4056548274(1)\\
Iwasaki          &7&.989846185(1) &9&.02297610282(2)  &9&.82968352735(1)\\
DBW2             &6&.346306461(1)  &7&.0639464328(2)  &7&.571097190(2) \\
\hline
\end{tabular}
\end{center}
\caption{The contribution $b_{S,P}$ to $\Lambda^{S,P}$ for various actions.}
\begin{center}
\label{values of bV,A}
\begin{tabular}{|c|r@{}l|r@{}l|r@{}l|}
\multicolumn{1}{c}{Action}&
\multicolumn{2}{c}{$\rho=0.6$} &
\multicolumn{2}{c}{$\rho=1.0$} &
\multicolumn{2}{c}{$\rho=1.4$} \\
\hline
Plaquette          &6&.3204575633(1)  &6&.33125072110(1)  &6&.34801586915(1)\\
Symanzik           &6&.3141428350(1)  &6&.32252999501(1)  &6&.3354222684(1)\\
TILW1 $\beta=8.60$ &6&.3107172748(1)  &6&.31780175061(1) &6&.32862040074(3)\\
TILW2 $\beta=8.45$ &6&.3105909672(1)  &6&.31762724546(3) &6&.32836942659(2)\\
TILW3 $\beta=8.30$ &6&.3104170577(1)  &6&.31738695411(2) &6&.32802384701(4)\\
TILW4 $\beta=8.20$ &6&.3103092794(1)  &6&.31723802434(3) &6&.32780966499(3)\\
TILW5 $\beta=8.10$ &6&.3101704948(1)  &6&.31704624086(2) &6&.32753386701(3)\\
TILW6 $\beta=8.00$ &6&.3099985210(1)  &6&.31680857518(1) &6&.32719209974(1)\\
Iwasaki         &6&.3057833598(1)  &6&.31095730979(1) &6&.31875098630(4)\\
DBW2            &6&.2975701799(1)  &6&.2996002928(1) &6&.30255799548(6)\\
\hline
\end{tabular}
\end{center}
\caption{The contribution $b_{V,A}$ to $\Lambda^{V,A}$ for various actions.}
\end{table}
\newpage
\vspace{5cm}
\begin{table}[ht]
\begin{center}
\label{values of bT}
\begin{tabular}{|c|r@{}l|r@{}l|r@{}l|}
\multicolumn{1}{c}{Action}&
\multicolumn{2}{c}{$\rho=0.6$} &
\multicolumn{2}{c}{$\rho=1.0$} &
\multicolumn{2}{c}{$\rho=1.4$} \\
\hline
Plaquette           &4&.7673340248(1)  &4&.35436733681(1)  &4&.0162353498(7)\\
Symanzik            &4&.86177933052(5) &4&.4732701854(1)   &4&.16189952383(6)\\
TILW1 $\beta=8.60$  &4&.9360280383(1) &4&.56582325908(4)  &4&.27391139850(6)\\
TILW2 $\beta=8.45$  &4&.93915112696(5) &4&.56969902605(5)  &4&.27857610740(2)\\
TILW3 $\beta=8.30$  &4&.94350212841(2) &4&.57509633367(4)  &4&.28506866099(4)\\
TILW4 $\beta=8.20$  &4&.94622847715(4) &4&.57847693279(6)  &4&.28913324224(3)\\
TILW5 $\beta=8.10$  &4&.9497748610(1)  &4&.58287278284(4) &4&.29441617825(3)\\
TILW6 $\beta=8.00$  &4&.95422443865(1)&4&.58838565488(6) &4&.30103785269(2)\\
Iwasaki          &5&.0777624156(2)  &4&.7402843779(1)  &4&.48177346976(3)\\
DBW2             &5&.6146580874(6)  &5&.378151584(1)   &5&.213044934(1)\\
\hline
\end{tabular}
\end{center}
\caption{The contribution $b_{T}$ to $\Lambda^{T}$ for various actions.}
\end{table}
\begin{table}[ht]
\begin{center}
\label{values of bnew}
\begin{tabular}{|c|r@{}l|r@{}l|r@{}l|}
\multicolumn{1}{c}{Action}&
\multicolumn{2}{c}{$\rho=0.6$} &
\multicolumn{2}{c}{$\rho=1.0$} &
\multicolumn{2}{c}{$\rho=1.4$} \\
\hline
Plaquette           &-15&.7863271766(1) &-15&.6250817532(1) &-15&.4858619898(1)\\
Symanzik            &-11&.9937499031(2) &-11&.8664388808(2) &-11&.7640746811(2) \\
TILW1 $\beta=8.60$  &-9&.7160722504(2)  &-9&.6095867774(2)  &-9&.5287008568(2)\\
TILW2 $\beta=8.45$  &-9&.6309328038(2)  &-9&.5252267216(1) &-9&.4451255929(1)\\
TILW3 $\beta=8.30$  &-9&.5135691580(2)  &-9&.4089372737(1)  &-9&.3299153795(1)\\
TILW4 $\beta=8.20$  &-9&.4407612246(2)  &-9&.3367955747(1)  &-9&.2584416246(1)\\
TILW5 $\beta=8.10$  &-9&.3468642074(4)  &-9&.2437578051(4)  &-9&.1662637020(4)\\
TILW6 $\beta=8.00$  &-9&.2303490445(1) &-9&.1283086414(1) &-9&.0518787415(1)\\
Iwasaki          &-6&.5572678902(2)  &-6&.4778880567(1) &-6&.4232808744(1)\\
DBW2             &0&.185440914(4)   &0&.211859722(4)     &0&.222852646(3)\\
\hline
\end{tabular}
\end{center}
\caption{The contribution $b$ to $\Lambda_{\mu\nu}$ for various actions.}
\end{table}
\begin{table}[ht]
\begin{center}
\label{values of bprimenew}
\begin{tabular}{|c|r@{}l|r@{}l|r@{}l|}
\multicolumn{1}{c}{Action}&
\multicolumn{2}{c}{$\rho=0.6$} &
\multicolumn{2}{c}{$\rho=1.0$} &
\multicolumn{2}{c}{$\rho=1.4$} \\
\hline
Plaquette           &0&.3497088259(1)  &0&.4340050439(1)   &0&.5267146951(2)\\
Symanzik            &0&.2719708888(4)  &0&.3367316099(2)   &0&.4080895543(1)\\
TILW1 $\beta=8.60$  &0&.2307195514(3)  &0&.2858316439(2)   &0&.3469216320(2)\\
TILW2 $\beta=8.45$  &0&.2290786112(2)  &0&.2838197180(1)   &0&.3445199656(2)\\
TILW3 $\beta=8.30$  &0&.2268014376(4)  &0&.2810293893(1)  &0&.3411911844(3)\\
TILW4 $\beta=8.20$  &0&.2253789601(2)  &0&.2792873534(1)   &0&.3391142249(3)\\
TILW5 $\beta=8.10$  &0&.2235391956(1)  &0&.2770353882(1)  &0&.3364306054(3)\\
TILW6 $\beta=8.00$  &0&.2212413920(1)  &0&.2742245511(1)  &0&.3330831698(2)\\
Iwasaki          &0&.1277382408(3)  &0&.1614096997(1)   &0&.2010432525(2)\\
DBW2             &-0&.116651374(1)  &-0&.118539399(1)   &-0&.111845323(1)\\
\hline
\end{tabular}
\end{center}
\caption{The contribution $b'$ to $\Lambda_{\mu\nu}$ for various actions.}
\end{table}

\vspace{5cm}

\begin{thebibliography}{99}
\bibitem{ref1}S. Capitani, \emph{Perturbative renormalization of the
  first two moments of non-singlet quark distributions with overlap
  fermions}, Nucl.Phys. B592 (2001) 183. 
\bibitem{DHK}T. DeGrand, A. Hasenfratz, T. Kovacs, \emph{Improving the
  chiral properties of lattice fermions}, Phys. Rev. D67 (2003) 054501.
\bibitem{ref10}C. Alexandrou, E. Follana, H. Panagopoulos and
  E. Vicari, \emph{One-loop renormalization of fermionic currents with the overlap-Dirac operator}, Nucl. Phys. B580 (2000) 394.
\bibitem{ref3}R. Horsley, H. Perlt, P. E. L. Rakow, G. Schierholz and A. Schiller, \emph{One-loop renormalization of quark bilinears for overlap fermions with improved gauge actions}, Nucl. Phys. B693 (2004) 3.
\bibitem{ref11}R. Horsley, H. Perlt, P. E. L. Rakow, G. Schierholz and
  A. Schiller, \emph{Renormalization of one-link quark operators for
    overlap fermions with L\"uscher-Weisz gauge action},
  Phys. Lett. B628 (2005) 66. 
\bibitem{ref2}M.Ioannou and H.Panagopoulos, \emph{Perturbative
  renormalization in parton distribution functions using improved
  actions}, PoS (LAT2005) 229.
\bibitem{ref4}K. Symanzik, \emph{Continuum limit and improved action in lattice theories. 1. Principles and phi**4 theory}, Nucl. Phys. B226 (1983) 187.
\bibitem{ref5}M. L\"uscher and P. Weisz, \emph{On-shell improved lattice gauge theories}, Commun. Math. Phys. 97 (1985) 59 [Erratum-ibid. 98 (1985) 433].
\bibitem{ref6}M. G. Alford, W. Dimm, G. P. Lepage, G. Hockney and P. B. Mackenzie, \emph{QCD on small computers}, Phys. Lett. B361 (1995) 87.
\bibitem{ref7}Y. Iwasaki, \emph{Renormalization group analysis of lattice theories and improved lattice action. 2. Four-dimensional nonabelian SU(N) gauge model},  UTHEP-118 (1983).
\bibitem{ref8}T. Takaishi, \emph{Heavy quark potential and effective actions on blocked configurations}, Phys. Rev. D54 (1996) 1050.
\bibitem{ref9}H. Neuberger, \emph{Exactly massless quarks on the
  lattice}, Phys. Lett. B417 (1998) 141; \emph{More about exactly
  massless quarks on the lattice},  Phys. Lett. B427 (1998) 353.
\bibitem{PS}A. Skouroupathis and H. Panagopoulos, \emph{Additive and
  multiplicative renormalization of topological charge with improved
  gluon/fermion actions: A test case for 3-loop vacuum calculations,
  using overlap or clover fermions}, 
Phys. Rev. D72 (2005) 094509.
\end{thebibliography}
\end{document}